\documentclass[12pt]{iopart}
\usepackage{widetext}
\usepackage{amssymb}
\makeatletter
\expandafter\let\csname equation*\endcsname\relax
\expandafter\let\csname endequation*\endcsname\relax
\makeatother
\usepackage{amsmath}

\usepackage{lineno}
\usepackage{cite}
\modulolinenumbers[5]
\usepackage[section]{placeins}
\usepackage{mathptmx}
\usepackage{graphicx}
\usepackage{tikz}
\usetikzlibrary{arrows, automata, positioning}

\usepackage{setspace}
\begin{document}
%\title[Finite State Logic]{Implementation of Finite State Logic Machines via the Dynamics of Atomic Systems}
\title[]{Implementation of Finite state logic machines via the dynamics of atomic systems}
\author{Dawit Hiluf Hailu\footnote{\ead{dhailu@bowiestate.edu}}}

\address{Department of Natural Sciences, Bowie State University, 14000 Jericho Park Road, Bowie, MD 20715-9465, USA}

\begin{abstract}
 
Following the success of Moore's predictions, we are approaching a limit in the miniaturization of semiconductors for computing materials. This has led to the exploration of various research paths to develop alternative computing paradigms, such as quantum computing, 3D transistors, molecular logic, and continuous logic. In this context, we propose a novel approach in which the dynamics of a two-level atom is used to execute classical Boolean logic operations. Unlike traditional combinational logic circuits, where the output depends solely on the input, we suggest a finite-state machine-like computing model, where the output is influenced by both the input and the system's initial state. The proposed mechanism leverages the dynamics of a two-level quantum state, with information encoded in observable quantities. These observables, the density matrix's population (diagonal) and coherence (off-diagonal) elements, were analyzed using the Liouville equation. The selection of observables within the Liouville space allows us to encode more variables. Although environmental noise may cause some loss of encoded information, fast computations can still be performed before it dissipates. In addition, logic operations can be read in parallel, enabling complex computations. This system can also be scaled to an N-level configuration.

\end{abstract}

%\begin{keyword}
%Two-level\sep CNOT gate \sep Magnus solution \sep population transfer \sep coherence
%%\MSC[2010] 00-01\sep  99-00
%\end{keyword}

%\end{frontmatter}

%\linenumbers

%%%%%%%%%%%%%%%%%%%%%%%%%%%%%%%%%%%%%%%%%%%%%%%%%
\section{Introduction}
\label{sec:intro}
%%%%%%%%%%%%%%%%%%%%%%%%%%%%%%%%%%%%%%%%%%%%%%%%%
Moore's law lies at the core of miniaturization and reduced power consumption of logic,  which is an observation made by Intel co-founder Gordon E Moore in that the number of transistors on an integrated circuit (IC) doubles about every two years \cite{moore2010cramming}. The ever-growing reliance of humans on technology demands fast computation, a large storage capacity,  and small and light computing machines that can be carried in a pocket. This, coupled with Moore's law, has led to the miniaturization of transistors to its edge. There is ongoing and widely growing interest among researchers in the search for alternatives. Quantum computing has the potential to deliver fast computations because it uses quantum bits, or qubits, to process information \cite{MichaelChuang2010}. A qubit is a unit of quantum information and the bit in quantum computing, whose counterpart in the classical bit, is $0s,1s$. Using quantum phenomenon,  we propose using coherences, the superposition of two states, to process our information.

The study of the dynamics of two-level systems has attracted the interest of researchers in different areas, ranging from nuclear magnetic resonance (NMR) \cite{Eberly1975} to Quantum Computers \cite{MichaelChuang2010}. Moreover, the dynamics of a two-level system mimics the logic operation of a Controlled-NOT (CNOT) gate \cite{hailu2019su2}. In this particular case, the Hamiltonian is closed under Lie algebra, and the solution can be obtained using Wei-Norman \cite{wei1963lie,fano1957description} and has been pursued by different authors \cite{dattoli1988evolution, altafiniuse}. As an application, the author of \cite {altafiniuse}, for example, has used the solution obtained via Wei-Norman for quantum computing. The solution obtained via Wei-Norman is exact; however, it is sometimes insightful to obtain analytical solutions to the system's dynamics. It is possible to use the knowledge of the eigenvalues and provide analytical solutions for the two-level system using the third-order Magnus expansion using the Sylvester formula \cite{hailu2019su2}. Although the focus is usually on an isolated system, it should be noted that including noise from the surrounding environment destroys coherence.

The molecular logic approach we adopt can take various forms, including those based on chemical, electrical, or optical signals, as outlined in the references \cite{beil2011logic, Collini:2013aa, Klein:2007aa, Klein:2009aa, Kompa:2001aa, klein2010ternary, levine2013realization, mol2011balanced, Remacle:2001aa, Remacle:2001ab, Remacle:2006aa, RemacLevine23, remacle2006all, remacle2008inter, FreschQDT}. In this paper, we focus on an all-optical input-output signal. This choice offers the advantage of enabling significantly faster computations compared to systems that rely on chemical or electrical signals. %However, despite the high speed of computation, it is also limited by the decay of the system's lifetime or dephasing, which can lead to the loss of stored information over time.
 
The paper is organized as follows. Section (\ref{sec:thePhyssystem}) introduces the system suitable for implementing the proposed machine, which involves rare-earth ions doped into a crystal. In Section (\ref{sec:thesystem}), we present a two-level system, describing its Hamiltonian and deriving the equation of motion. Section (\ref{sec: Analytical}) outlines the application of Sylvester's formula to obtain the analytical solution to the equation of motion. Section (\ref{sec: results}) provides numerical solutions. In Section (\ref{sec: parity}), we explore how physical systems can emulate a finite-state machine, with a focus on a parity checker. Section (\ref{sec: scalability}) examines the scalability of the proposed machine to an N-level system. Section (\ref{sec: LSM}) offers a detailed description of the new machine, while section (\ref{sec:compare}) compares it with a finite-state machine. Section (\ref{sec:future}) discusses future directions, particularly efforts to mitigate noise. Finally, Section (\ref{sec: conclusion}) summarizes our findings and offers perspectives for future research.
%%%%%%%%%%%%%%%%%%%%%%%%%%%%%%%%%%%%%%%%%%%%%%%%%%%%%%%%%%%%%%%%%%%%%%%%%%%%%%%%  
\section{Potential Material for Implementation of the Proposed Machine.}  
\label{sec:thePhyssystem}
\begin{figure}[htbp]
\begin{center}
(a) \includegraphics[width=2.5 in]{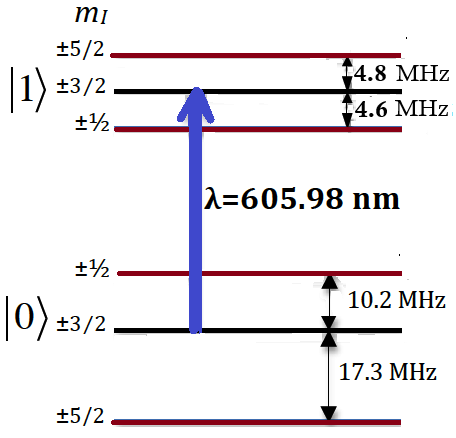}
(b) \includegraphics[width=2.3 in]{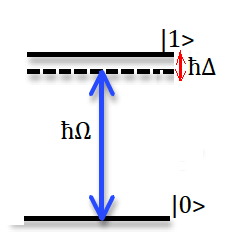}
\caption{(Color online) (a) Energy level diagram of Pr$^{3+}$:Y$_2$SiO$_5$ (b) The corresponding two-level system with detuning $\hbar\Delta$ and coupling pulse $\hbar\Omega\left(t\right)$.}
\label{thsyevel}
\end{center}
\end{figure}

%%%%%%%%%%%%%%%%%%%%%%%%%%%%%%%%%%%%%%%%%%%%%%%%%%%%%%%%%%%%%%%%%%%%%%%%%%%%%%%%  
In this section, we present the candidate material suited for implementing the proposed logic, which consists of ions of rare earth ion doped into the crystal. Recent studies by the authors of \cite{klein2008experimental, beil2008electromagnetically, Klein:2007aa} have demonstrated the successful use of population transfer through stimulated Raman adiabatic passage (STIRAP) with Praseodymium ions (Pr$^{3+}$) doped into a Y$_2$SiO$_5$ crystal. Rare-earth ions in an inorganic crystal exhibit optical transitions that differ from those in isolated ions due to the interaction with the crystal field. The energy level structure of Pr$^{3+}$:Y$_2$SiO$_5$, hereafter referred to as Pr:YSO, is shown in Fig.(\ref{thsyevel}a). These optical transitions in Pr:YSO are characterized by narrow homogeneous linewidth, typically in the range of a few kHz. However, crystal impurities lead to variations in the crystal field, causing inhomogeneous broadening of the optical transitions by several GHz. This broadening means that a single-frequency laser will address multiple ions within the broadened spectral line. The benefit of using rare-earth ions in this context is their ability to mitigate the loss or destruction of information before the system performs its computation. The long decoherence times and narrow linewidth help preserve the quantum states for extended periods, reducing the impact of environmental noise and imperfections, which are crucial for reliable computation and information transfer.
%\graphicspath{{Images//}}
Rare-earth ions, such as praseodymium ions (Pr$^{3+}$) in Y$_2$SiO$_5$, exhibit exceptionally long decoherence times, often ranging from milliseconds to seconds under optimal conditions. In high-quality crystals, these coherence times can exceed 1 millisecond, making them particularly suitable for the storage and processing of quantum information \cite{klein2008experimental}. These extended decoherence times enable stable quantum memories, allowing information to be stored for long periods without significant loss of coherence. This feature is especially valuable in quantum communication and computation, where maintaining quantum states over time is critical. Moreover, the narrow homogeneous linewidth of the optical transitions in these systems (on the order of a few kHz) further minimizes the effects of environmental disturbances, enhancing the stability and reliability of quantum operations \cite{beil2008electromagnetically}. These characteristics position rare-earth-ion-doped solids as a promising platform for implementing the proposed logic machine.

Mitigating information loss and protecting against decoherence is a vast and active area of ongoing research \cite{Viola2005, Lidar2013, ShorDeco, SteaneErrorQM, BealeQMEr}. As with quantum computation, information stored in coherences is susceptible to disruption due to environmental interactions. However, we argue that the proposed logic, which utilizes optical signals, processes information quickly enough to avoid significant decoherence. Specifically, the computational steps in this framework are designed to be completed well within the coherence time, ensuring that the quantum states remain preserved throughout the computation. In rare-earth-ion-based systems, decoherence times typically range from milliseconds to seconds, depending on the material and experimental conditions \cite{klein2008experimental}. In contrast, the computation timescale for the proposed logic is much shorter than these decoherence times. This rapid processing ensures that operations are completed before decoherence can substantially affect stored information. Therefore, the approach outlined here leverages the extended coherence times of rare-earth ions, facilitating reliable quantum computation and information transfer within the available coherence window \cite{Beil:2009aa, Vitanov:2001aa, McAuslan}.

It is worth stressing that the long decoherence times exhibited by rare-earth ions in solid-state hosts are one of their most significant advantages, particularly in quantum information processing and storage. Decoherence refers to the loss of quantum coherence between the different states of a quantum system because of its interaction with the environment. In many quantum systems, decoherence occurs on timescales that are too short for practical quantum computation or information storage. However, rare earth ions, because of their unique properties, can maintain coherence for much longer periods, making them ideal candidates for our case.

%In our model, we treat the system as a two-level quantum system, as shown in Fig. (\ref{thsyevel}b), with states $|0\rangle$ and $|1\rangle$, corresponding to energies $\hbar \omega_0$ and $\hbar \omega_1$, respectively. These two levels are coupled by laser pulses that drive the transition between $|0\rangle$ and $|1\rangle$\cite{vitanov1999creation, shore2011manipulating}.

%%%%%%%%%%%%%%%%%%%%%%%%%%%%%%%%%%%%%%%%%%%%%%%%%
\section{The two-level system}
\label{sec:thesystem}
%%%%%%%%%%%%%%%%%%%%%%%%%%%%%%%%%%%%%%%%%%%%%%%%%
The considered system is shown in Figure~(\ref{thsyevel}b), which we assume to have a two-level system with ground state $|0\rangle$ and excited states $|1\rangle$. Suppose that the two levels are coupled by laser pulses $\Omega \left(t\right)$, which drive the transition between $|0\rangle$ and $|1\rangle$\cite{vitanov1999creation, shore2011manipulating}. The laser is off by detuning \(\Delta\), which is the difference between the laser frequency \(\omega_L\) and the Bohr frequency \(\omega_0\). In a two-level quantum system, the Bohr frequency \(\omega_0\) represents the natural transition frequency between the two energy levels and is given by \(\omega_0 = \frac{E_1 - E_0}{\hbar}\), where \(E_0\) and \(E_1\) are the energy levels and \(\hbar\) is the reduced Planck constant. The laser frequency \(\omega_L\) is the frequency at which the laser light is applied to the system. The detuning \(\Delta\) is defined as \(\Delta = \omega_L - \omega_0\), indicating how far the laser frequency deviates from the resonance frequency of the transition. When \(\Delta\) is zero, the laser frequency is precisely aligned with the Bohr frequency, putting the system in resonance \cite{Eberly1975}. %Conversely, when \(\Delta\) is nonzero, the laser frequency is either above (\(\Delta > 0\)) or below (\(\Delta < 0\)) the Bohr frequency, resulting in off-resonance conditions \cite{Eberly1975}. %By setting \(\Delta\) to a value significantly different from zero, the laser frequency deviates from the Bohr frequency, thereby diminishing its effect on the system and effectively turning off the laser’s influence on the transition.

We consider an adiabatic population transfer, which involves gradually transferring the population of a quantum system from one energy state to another by slowly varying external parameters. This process relies on the principle that if the perturbation is applied slowly relative to the system's internal dynamics, the system remains in its instantaneous eigenstate \cite{Bergmann2001,shore2011manipulating,breuer2007theory}. However, we applied a weak pulse (corresponding to a pulse area of $\pi/2$) to avoid transferring the entire population from $|0\rangle$ to $|1\rangle$, as would have been the case when using a pulse area of $\pi$. Using the pulse area protocol $\pi$, the populations could completely transfer from the ground state $|0\rangle$ to the excited states $|1\rangle$ \cite{Bergmann2001}.

The time evolution of an $N$-level atomic system is described by the density matrix $\hat{\rho}$. This matrix evolves following the Liouville equation \cite{levine2011quantum, cohen1991quantum}, which dictates the changes in the state of the system over time. The Liouville equation incorporates both the coherent dynamics of the system and its interactions with the surrounding environment, offering a thorough framework for analyzing the system's behavior.
\begin{equation}
\begin{aligned}
-i\hbar\frac{\partial}{\partial t}\hat\rho=&\left[\hat\rho,\hat H\right]
\label{Liouville}
\end{aligned}
\end{equation}

Using the projector operator $\hat{O}_{mn} = |m\rangle\langle n|$, we find the expectation value as $\langle\hat{O}_{mn}(t)\rangle = \text{Tr}(\hat{O}_{mn} \hat{\rho}(t)) = \rho_{mn}(t)$. This indicates that the dynamics of the system can be explored through the operator $\hat{O}_{mn}$. In this section, we will derive the equation of motion for the three-dimensional coherence vector using the Alhassid-Levine formalism \cite{alhassid1977entropy}. Alternatively, the equations of motion can also be derived using the Hioe-Eberly formalism \cite{HIOE:1981aa, hioe1982nonlinear, hioe1983dynamic}. 

The Hamiltonian under Rotating Wave Approximation (RWA) and in the interaction picture is known as \cite{shore2011manipulating, Shore:2008aa}.
\begin{equation}
\begin{aligned}
\hat H \left(t\right)=\frac{\hbar}{2}\begin{pmatrix}
0 & \Omega  \left(t\right)\\
\Omega \left(t\right) & 2\Delta \\
 \end{pmatrix} 
  \end{aligned}
\end{equation}
One can also express the Hamiltonian in terms of the projection operators $\hat{O}_{mn} = |m\rangle\langle n|$, i.e., $|0\rangle\langle1|$, $|1\rangle\langle0|$, and $|1\rangle\langle1|$ as:
\begin{equation}
\begin{aligned}
\hat{H}(t) = \frac{\hbar}{2} \Omega(t) \left( |0\rangle\langle1| + |1\rangle\langle0| \right) + \hbar\Delta |1\rangle\langle1|
\end{aligned}
\label{Hamil_op}
\end{equation}
%\textbf{Solution}: To begin, we construct the coherence vector using generators that form a closed Lie algebra, as discussed in \cite{hailu2019su2, DHCNOT} and related references. For clarity and future reference, we restate here the relationship between the generators and the components of the coherence vectors as
We use Pauli matrices as our generators, as outlined in \cite{MichaelChuang2010}. It is important to note that these matrices can be expressed as linear combinations of projector operators \(\hat{O}_{mn} = |m\rangle\langle n|\).
\begin{equation}
\begin{aligned}
\hat s_1=& |0\rangle\langle1| + |1\rangle\langle0|,\\
\hat s_2=&-i(|0\rangle\langle1| - |1\rangle\langle0|),\\
\hat s_3=&|0\rangle\langle0| - |1\rangle\langle1|
\label{sigma}
\end{aligned}  
\end{equation}  
 
%\begin{equation}
%\begin{aligned}
% \hat s_1=\begin{pmatrix}
%0 & 1\\
%1 & 0 \end{pmatrix},       & & \hat s_2=\begin{pmatrix}
%0 & -i\\
%i & 0 \end{pmatrix} ,      & & \hat s_3=\begin{pmatrix}
%1 & 0\\
%0 & -1 
% \end{pmatrix}  
% \end{aligned}
% \label{sigma}
%\end{equation}
In this regard, we expand the density matrix and the Hamiltonian \cite{Hioe1981,Fresch:2013aa,Hiluf_2018}.
 \begin{equation}
\begin{aligned}
\hat\rho\left(t\right)=&\frac{\hat I}{2}+\frac{1}{2}\sum_{j=1}^{3} S_j\left(t\right)\hat s_j\\
\hat H\left(t\right)=&\frac{\hbar}{2}\left[\left(\sum_{k}^2\omega_k\right)\hat I+\sum_{j=1}^{3} h_j\left(t\right)\hat s_j\right]
\end{aligned}
\end{equation}
where $\hbar\omega_k$ is the energy of level $k$ ($\omega_k$ is the frequency of level $k$) and $\hat I$ is the identity operator. The coefficients $S_j\left(t\right)$ and $h\left(t\right)$ are given by \cite{Hioe1981}
\begin{subequations}
\begin{align}
S_j\left(t\right)=&Tr\left(\hat\rho\left(t\right)\hat s_j\right)\label{expectS}\\
h_j\left(t\right)=&\frac{1}{\hbar}Tr\left(\hat H\left(t\right)\hat s_j\right)\label{Torque}
\end{align}
\end{subequations}
the generators $\hat s_j$ are the Pauli matrices and has the following properties \cite{Hioe1981}
\begin{equation}
\begin{aligned}
Tr\left(\hat s_i\hat s_j\right)=& 2\delta_{ij}\\
\left[\hat s_{i}, \hat s_{j}\right]=&2i \epsilon_{ijk} \hat s_k
\end{aligned}
\end{equation}
where $\delta_{ij}$ is the Kronecker delta and $\epsilon_{ijk}$ is the Levi-Civita which are antisymmetric in all three indices.

Based on this, we obtain the equation of motion for the coherence vector and follow the same procedure as in \cite{Hioe1981, AlhassidLevine}. If we now form a vector $\vec S=\left(S_1\left(t\right), S_2\left(t\right), S_3\left(t\right)\right)^T$ whose elements are the expectation value of the Pauli matrices as given by Eq.\eqref{expectS}, we can readily obtain the equation of motion as follows:
\begin{equation}
\begin{aligned}
\frac{d}{dt}\vec{S}(t)=&g(t)\vec S(t),      &     \vec S \left(t_0\right) =S_0
\label{sdot}
\end{aligned}
\end{equation}
 where $g$ is given by 
\begin{equation}
\begin{aligned}
g(t)=&
 \begin{pmatrix}
  0 & \Delta  & 0 \\
  -\Delta & 0 & -\Omega\left(t\right) \\
  0 & \Omega\left(t\right) & 0
 \end{pmatrix}
\end{aligned}
\label{matrixg}
\end{equation}
%%%%%%%%%%%%%%%%%%%%%%%%%%%%%%%%%%%%%%%%%%%%%%%%%%%%%%%%%%%%%%%%%%%%%%%%%%%%%%%%
\section{Analytical Solution}
\label{sec: Analytical}
%%%%%%%%%%%%%%%%%%%%%%%%%%%%%%%%%%%%%%%%%%%%%%%%%%%%%%%%%%%%%%%%%%%%%%%%%%%%%%%%
Although the exact equation of motion, Eq. \eqref{sdot}, can be solved numerically; we are interested in finding an analytical solution to use for the operation of finite state machines (FSM). %We will solve Eq.~\eqref{sdot} numerically. %To show the correspondence of how the solution emulates the proposed finite state machine (FSM) with an analytical solution. The inclusion of an analytical solution is essential for establishing a link between the coherence vector's solution and the finite state machine (FSM), which is a computational model whose output dependes on both the input and the state the machine is in. An FSM consists of a finite number of states, transitions between these states, and actions, and it operates based on input signals, transitioning from one state to another according to predefined rules. By establishing this connection, we can relate the input generated by the system's dynamics to the corresponding states of the FSM. This relationship enables us to better understand how the evolution of the coherence vector mirrors the state transitions within the machine. This analysis provides insight into the dynamic system's behavior about the discrete states of the machine, offering a comprehensive perspective on how continuous system parameters correspond to discrete state changes 
To demonstrate how the solution mimics the proposed finite state machine (FSM) using an analytical approach, it makes sense to include an analytical solution. This is necessary to establish a connection between the coherence vector's solution and the FSM, a computational model where the output depends on both the input and the current state. An FSM is characterized by a finite number of states, transitions among these states, and actions that operate based on input signals and transitions between states according to predefined rules. By linking the system's input dynamics to the FSM states, we can understand how the coherence vector's evolution reflects the FSM's state transitions. This analysis offers insights into the behavior of the dynamic system to the discrete states of the FSM, providing a comprehensive view of how the parameters of the continuous system align with the changes in the discrete state\cite{parallelKsenia,dynamicsKsenia,DnstyQMDynamicsKsenia,Hiluf_2018}. %; in search of a better approximation, the Magnus solution must include the $3^{rd}$ order expansion.   
%%%%%%%%%%%%%%%%%%%%%%%%%%%%%%%%%%%%%%%%%%%%%%%%%%%%%%%%%%%%%%%%%%%%%%%%%%%%%%%%
%\section{Magnus Approximations}
%\label{sec: Magnus}
%%%%%%%%%%%%%%%%%%%%%%%%%%%%%%%%%%%%%%%%%%%%%%%%%%%%%%%%%%%%%%%%%%%%%%%%%%%%%%%%

Several methods, including the Magnus approximation and the spectral decomposition, can be used to solve Eq.~\eqref{sdot} analytically, and the methods are reviewed in \cite{19dubiousways,magnus1954exponential,altafiniuse,Aravind:86}. %The Magnus expansion, named after Wilhelm Magnus, provides an exponential representation of the solution of a first-order linear homogeneous equation for a linear operator. 
Given a $3\times 3$ coefficient matrix $g\left(t\right)$, we want to solve the initial value problem associated with the linear ordinary differential equation, which in our case is the equation of motion for the coherence vectors given by Eq.~\eqref{sdot}, along with its initial condition. The solution can be written as
\begin{equation}
\begin{aligned}
\vec S\left(t\right)=R\left(t_0,t\right)\vec S\left(t_0\right)
\end{aligned}
\label{Rsol}
\end{equation}
%The solution can be written as $\vec S\left(t\right)=R\left(t,t_0\right)\vec S\left(t_0\right)$;
%\begin{equation}
%\begin{aligned}
%\vec S\left(t\right)=&R\left(t,t_0\right)\vec S\left(t_0\right)
%\end{aligned}
%\label{su2gsol}
%\end{equation}
where $\vec S$ is a 3-dimensional column vector and $R$ is a $3\times3$ superevolution matrix. This solution provides insight into the final state of the expectation values of the generators. Starting with the initial state of these expectation values, the application of an input, such as a laser field, induces changes in the population of the two states, consequently altering the final state of the expectation values of the generators.

%%%%%%%%%%%%%%%%%%%%%%%%%%%%%%%%%%%%%%%%%%%%%%%%%%%%%%%%%%%%%%
%%%%%%%%%%%%%%%%%%%%%%%%%%%%%%%%%%%%%%%%%%%%%%%%%%%%%%%%%%%%%%%%%%%%%%%%%%%%%%%%
\subsection{Sylvester's Formula} 
%%%%%%%%%%%%%%%%%%%%%%%%%%%%%%%%%%%%%%%%%%%%%%%%%%%%%%%%%%%%%%%%%%%%%%%%%%%%%%%%
%The approach outlined in subsection (\ref{adiab}) assumes that the Hamiltonian of the system commutes with itself at different times. However, 
In the case of ordinary differential equations (ODEs) with time-varying coefficients, such as the one under consideration here, that is, Eq.~\eqref{sdot}, the general solution is approximated by

\begin{equation}
\begin{aligned}
\vec{S}\left(t\right)=& \mathcal{T}\{e^{\int_{t_0}^t g\left(t'\right) dt'}\}\vec S \left(t_0\right), 
\end{aligned}
\label{soltmordg}
\end{equation}
where $ \mathcal{T}$ denotes time-ordering,
\begin{equation}
\begin{aligned}
\mathcal{T}\{e^{\int_{t_0}^t g\left(t'\right) dt'}\}\equiv & \sum_{n=0}^\infty\frac{1}{n!}\int_{t_0}^t\ldots\int_{t_0}^t\mathcal{T}\{g\left(t_1'\right)\ldots g\left(t_n'\right)\} ~dt_1'\ldots dt_n'\\
=&\sum_{n=0}^\infty\int_{t_0}^t dt_1'\ldots\int_0^{t'_{n-1}}dt'_n~ g\left(t_1'\right)\ldots g\left(t_n'\right)
\end{aligned}
\end{equation}

Assuming the matrices commute at different times, that is, the commutator $[g(t_1), g(t_2)] = 0 $ for all $ t_1, t_2 $, the time-ordered expression takes the form $ e^{\int_{t_0}^t g(t') \, dt'} $.  The evaluation of such an exponential has been extensively studied \cite{wilcox1967exponential,wei1963lie,wei1964global,magnus1954exponential,Suzuki1976}. In this section, we utilize the Sylvester formula to derive analytical solutions for coupled differential equations.

%%%%%%%%%%%%%%%%%%%%%%%%%%%%%%%%%%%%%%%%%%%%%%%%%%%%%%%%%%%%%%%%%%%%%%%%%%%%%%%%
%\subsubsection{Sylvester's formula for distinct eigenvalues} 
%%%%%%%%%%%%%%%%%%%%%%%%%%%%%%%%%%%%%%%%%%%%%%%%%%%%%%%%%%%%%%%%%%%%%%%%%%%%%%%%
%Sylvester's formula is a widely recognized tool for solving exponential equations. 
Given an $ n \times n $ coefficient matrix $g(t) $, our objective is to solve the initial value problem related to the linear ordinary differential equation that governs the equation of motion of coherence vectors (see Eq.~\eqref{sdot}). %For clarity and convenience, we rewrite this equation along with its initial condition as
%\begin{equation}
%\begin{aligned}
%\frac{d }{dt}\vec{S}\left(t\right)=&g\left(t\right)\vec S \left(t\right), &     \vec S %\left(t_0\right) =S_0
%\end{aligned}
%\end{equation}
The solution, assuming the matrix $g$ is commutable with itself at different times,  can be expressed as 
\begin{subequations}
\begin{align}
\vec{S}\left(t\right)=&e^{\int_{t_0}^t g\left(t_1\right) dt_1}\vec S \left(t_0\right)\\
\vec{S}\left(t\right)=&e^{G\left(t\right)}\vec S \left(t_0\right)\label{expsoln}
\end{align}
\end{subequations}

Where $G(t) = \int_{t_0}^t g(t_1) \, dt_1 $, over the time interval $[t_0,t]$. Here, we assume that all the entries $ g_{ij} $ of the matrix $g $ are integrable functions. Consequently, we define the integral of the matrix as the matrix whose entries are integrals of the corresponding elements of $g $, expressed mathematically as $ \int g(t) \, dt:= \left( \int g_{ij}(t) \, dt \right) $. 

Sylvester's formula provides a method for computing the exponential of a matrix using only its eigenvalues \cite{moler2003nineteen,tarantola2006elements}. %Therefore, for an~ $ n\times n $ matrix, in our case for $N$ level system, we point out that $n=N^2-1$, the solution to its exponential can be obtained by finding its eigenvalues and applying Sylvester's formula. Using this approach, we can express the exponent in equation \eqref{expsoln} as
For an $ n \times n $ matrix, and specifically for an $ N $-level system where $ n = N^2 - 1 $ is considered in this paper, the solution to the matrix exponential can be obtained by computing its eigenvalues and applying Sylvester's formula. Using this approach, we can express the exponent in equation \eqref{expsoln} as
\begin{equation}
\begin{aligned}
e^{G\left(t\right)}=&\sum_{j=1}^{N^2-1}e^{\lambda_j}\prod_{j\neq k=1}^{N^2-1}\frac{G\left(t\right)-\lambda_j  I}{\lambda_k-\lambda_j}
\end{aligned}
\label{expSylv}
\end{equation}
Where $ \lambda_j $ denotes the eigenvalues of $G(t) $, $ I $ represents the identity matrix, and $ e^{\lambda_j} $ signifies the exponent of the eigenvalues.

Let us apply Sylvester's formula to derive the analytical solution of the coherence vector, or Bloch vector, for a two-level system. The eigenvalues of $g$ are readily obtained as $\{0, -\sqrt{-\Delta'^2 - \Omega'^2}, \sqrt{-\Delta'^2 - \Omega'^2}\}$, where $\Delta'= \int_{t_0}^t \Delta \, dt'_1 $ and $\Omega'= \int_{t_0}^t \Omega \, dt'_1 $ are the integrated values of $\Delta$ and $\Omega$, respectively. 
By substituting the initial values of the coherence vector, where the population is initially prepared in the ground state $|0\rangle$, $\vec{S}(t_0) = (0, 0, 1)^T$, which follows from the definition of the vector $\vec{S} = (\hat s_1, \hat s_2, \hat s_3)^T$, we find that the solution can be expressed as
\begin{equation}
\begin{aligned}
S\left(t\right)=~&e^{G\left(t\right)}\vec S \left(t_0\right)=\Big[\sum_{j=1}^{3}e^{\lambda_j}\prod_{j\neq k=1}^{3}\frac{G\left(t\right)-\lambda_j  I}{\lambda_k-\lambda_j}\Big]\cdot\vec S \left(t_0\right)\\
=&\Big[e^{\lambda_1}\frac{(G\left(t\right)-\lambda_2I_{3\times3})(G\left(t\right)-\lambda_3I_{3\times3})}{(\lambda_1-{\lambda_2})(\lambda_1-\lambda_3)}+e^{\lambda_2}\frac{(G\left(t\right)-\lambda_1I_{3\times3})(G\left(t\right)-\lambda_3I_{3\times3})}{(\lambda_2-\lambda_1)(\lambda_2-\lambda_3)}\\
&+e^{\lambda_3}\frac{(G\left(t\right)-\lambda_1I_{3\times3})(G\left(t\right)-\lambda_2I_{3\times3})}{(\lambda_3-\lambda_1)(\lambda_3-\lambda_2)}\Big]\cdot\vec S \left(t_0\right).
\end{aligned}
\label{slvfrm}
\end{equation}
Introducing $\zeta = \sqrt{\Delta'^2 + \Omega'^2}$ and recalling the eigenvalues to be $\lambda_1 = 0$, $\lambda_2 = -\imath \zeta$, and $\lambda_3 = \imath \zeta$, we substitute these values in Eq.\eqref{slvfrm}. After performing some algebra, the solution takes the following form:
\begin{equation}
\begin{aligned}
\vec{S}\left(t\right)=&
 \begin{pmatrix}
 \frac{\Delta'\Omega'}{\zeta^2}\big(1-\cos\zeta\big)\\
  \frac{\Omega'}{\zeta}\sin\zeta\\
 -\frac{\Delta'^2}{\zeta^2}-\frac{\Omega'^2}{\zeta^2}\cos\zeta
 \end{pmatrix}
\end{aligned}
\label{sylv2l}
\end{equation}
The solution given by Eq.~\eqref{sylv2l} is valid when the system is initially prepared in the ground state. However, when the system is prepared in the excited state \( |1\rangle \), the initial condition of the system is given by \(\vec{S}(t_0) = (0, 0, -1)^T\). In this scenario, the solution can be expressed in the following form.  

\begin{equation}
\begin{aligned}
\vec{S}\left(t\right) = -&
\begin{pmatrix}
\frac{\Delta'\Omega'}{\zeta^2}\big(1-\cos\zeta\big)\\
\frac{\Omega'}{\zeta}\sin\zeta\\
-\frac{\Delta'^2}{\zeta^2}-\frac{\Omega'^2}{\zeta^2}\cos\zeta
\end{pmatrix}.
\end{aligned}
\label{sylv2l1}
\end{equation}  

It is essential to note that this solution is the negative of the solution obtained in Eq.~\eqref{sylv2l}. This inversion reflects the change in the system's initial condition and its impact on the resulting dynamics. The sign change is significant because it emphasizes how the initial state of the system influences the trajectory of its evolution. The negative sign in the solution underscores the versatility of the system, showing that the dynamics can reverse or invert depending on the initial conditions. This property is critical in applications where state manipulation and control are required, such as in quantum computation or decision-making algorithms. Additionally, the relationship between the solutions for different initial conditions demonstrates the coherence and symmetry inherent in the system's mathematical framework, making it a compelling platform for emulating FSM behavior in quantum-mechanical settings. 

Furthermore, the input applied to the system is the pumping laser, characterized by the Rabi frequency \(\Omega'\). The presence of \(\Omega'\) directly influences the evolution of the system, as it determines the strength of the interaction between the system and the laser field. Consequently, \(\Omega'\) also defines the parameter \(\zeta\), which encapsulates the cumulative effect of the laser field on the system over time. These parameters—\(\Omega'\) and \(\zeta\)—serve as external inputs driving the system's dynamics.  

In this context, Eq.~\eqref{Rsol} further elucidates that the future state of the system is governed by both its current state and the external input applied to it. This interplay between the current state and the input highlights the system's ability to emulate the behavior of a finite state machine (FSM). In an FSM, the output at any given time depends on the combination of the current state and the input. Similarly, the evolution of the quantum system depends on these two factors, demonstrating its ability to perform state-dependent operations that mirror computational logic.  

Here, the coupling between the initial state \(\vec{S}(t_0)\) and the external inputs \(\Omega'\) and \(\zeta\) determines the trajectory of the system. This interdependence highlights a core feature of the system: the output is a result of both the applied inputs and the system's initial state.  
This behavior is closely aligned with the principles of a finite-state machine (FSM). In an FSM, the output depends on the current state and the input. Similarly, in this quantum system, the external input (the pumping laser) and the initial state jointly dictate the system's evolution and output. The parameter \(\Omega'\) governs the rate and strength of the transitions between states, while \(\zeta\) encapsulates the accumulated effect of the laser over time. Together, these parameters act as the "control signals" of the quantum FSM, steering its dynamics, and determining its eventual state.  

%%%%%%%%%%%%%%%%%%%%%%%%%%%%%%%%%%%%%%%%%%%%%%%%%%%%%%%%%%%%%%%%%%%%%%%%%%%%%%%%%
%%%%%%%%%%%%%%%%%%%%%%%%%%%%%%%%%%%%%%%%%%%%%%%%%%%%%%%%%%%%%%
\section{Results and discussion}
\label{sec: results}
%%%%%%%%%%%%%%%%%%%%%%%%%%%%%%%%%%%%%%%%%%%%%%%%%%%%%%%%%%%%%%
Starting from the expression given in Eq.~\eqref{sylv2l} and setting $\Delta' = 0$ to simplify the equation, we find that $\zeta$ simplifies to $\Omega'$ when $\Delta' = 0$. Therefore, the vector $\vec{S}(t)$ under this condition becomes:
\begin{equation}
\begin{aligned}
\vec{S}(t) = 
\begin{pmatrix}
0 \\
\sin\Omega' \\
-\cos\Omega'
\end{pmatrix}
\end{aligned}
\label{sylv2ld0}
\end{equation}

This expression relates directly to Rabi oscillations, with $\Omega'$ representing the Rabi frequency that characterizes the interaction strength between the two-level system and the external driving field. In this context, $\Omega'$ dictates the oscillatory behavior of the system. The component $S_1$ is zero, indicating that there is no oscillation. The component $S_2$, $\sin\Omega'$, describes the coherence between the two states, oscillating sinusoidally with the Rabi frequency. The component $S_3$, $-\cos\Omega'$, represents the oscillatory variation in the population difference between the states. Collectively, these components depict how both the coherence and population differences oscillate with the Rabi frequency $\Omega'$, highlighting the system's response to the external field.

%Now that the relevant equations necessary to proceed are complete, let us briefly compare the analytical and numerical solutions.
For the case at hand, it is worth noting that the pulse profile we used prevents the observation of Rabi oscillations. Instead, at the end of the pulse interaction, we achieve a superposition of states, which effectively redistributes the population between states $|0\rangle$ and $|1\rangle$. To see the Rabi oscillation, wherein the population fluctuates between states $|0\rangle$ and $|1\rangle$, one needs to use a laser whose pulse area is an integral multiple of $\pi$.  
Numerical solutions are obtained for the case where $\Delta=0$, and the pulse profile is $\Omega\left(t\right)=\Omega_0 e^{\frac{-(t-\tau)^2}{\sigma^2}}.$ Here, $\Omega_0$  is the peak Rabi frequency,  $\tau$ is the time at which the pulse is centered and $\sigma$ controls the width of the pulse. The parameters are chosen in reduced units, meaning that they are scaled by natural or characteristic units of the system to simplify calculations and interpretations. 
\begin{widetext}
\graphicspath{{Images//}}
\begin{figure*}[htp]
\centering
(i)\includegraphics[width=3.0 in]{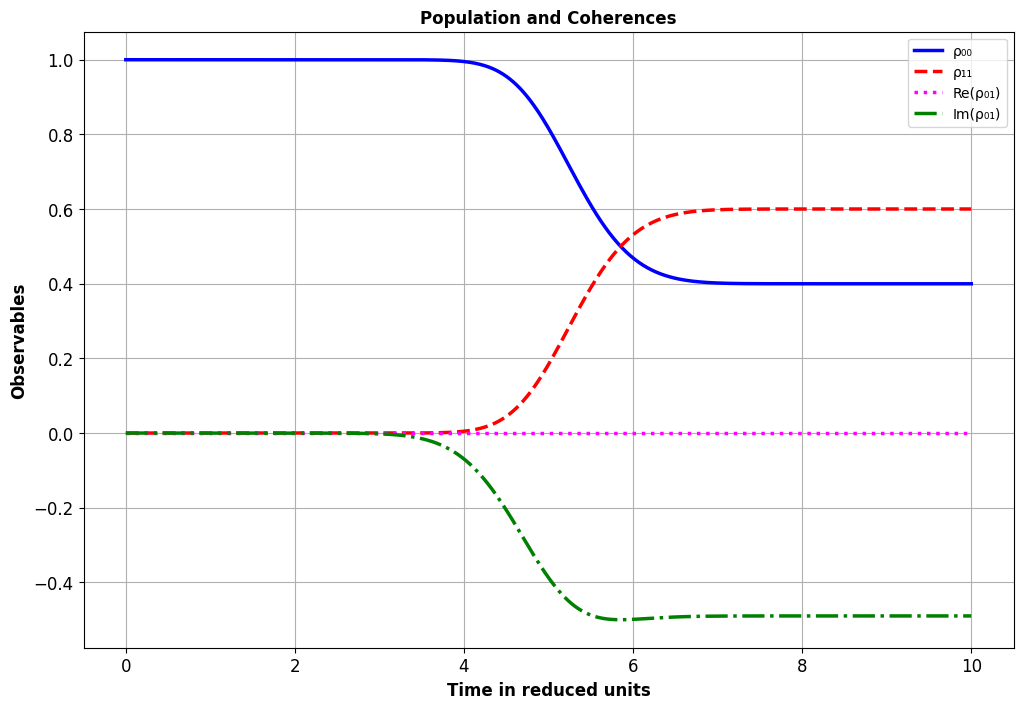}%\hfill
(ii)\includegraphics[width=3.0 in]{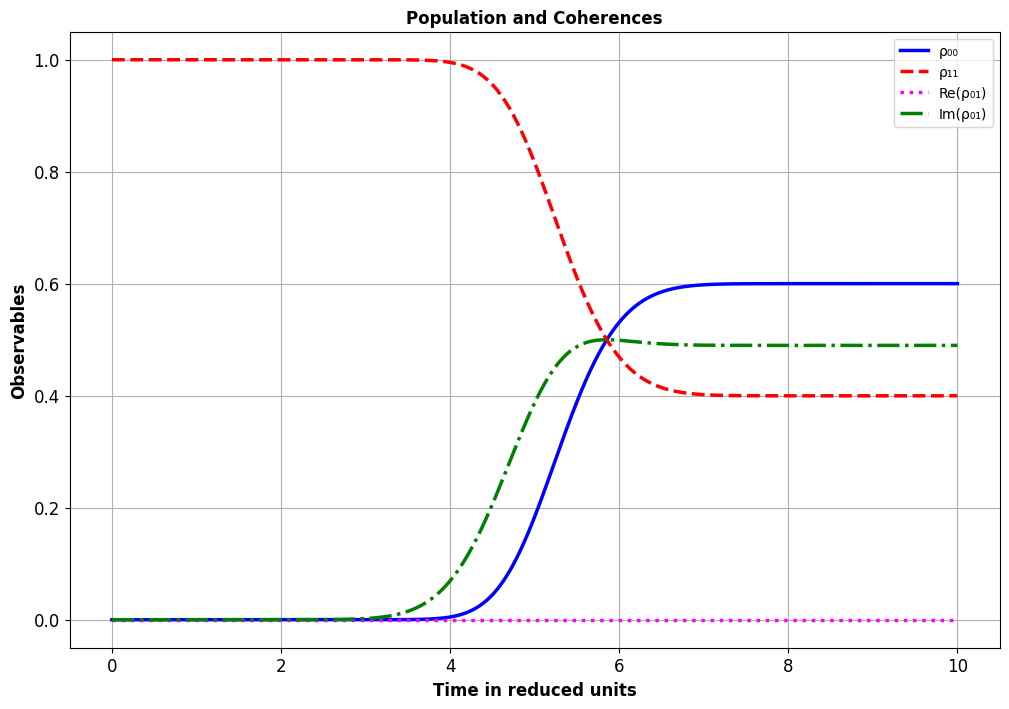}%\hfill
%(c)\includegraphics[width=.3\textwidth]{Pulse}%\hfill
%(d)\includegraphics[width=.3\textwidth]{Obser_S01}
\caption{(Color online) Numerical solutions for the case when $\Delta=0$. (i) When the population is initially prepared to be in the ground state $|0\rangle.$ (ii) When the population is initially prepared to be in the excited state $|1\rangle.$ Parameters in reduced units,  $\Omega_0 = 1.0$,  $\tau = 5.0$, and $ \sigma = 1.0$. Key: Solid line (blue) represents $\rho_{00}$, dashed line (red) represents $\rho_{11}$, dashed-dot line (green) denotes the real part of $\rho_{01}$, and dotted line (magenta) indicates the Imaginary part of $\rho_{01}$. Note that \(\text{Re}(\rho_{01})\) refers to the coherence in phase, while \(\text{Im}(\rho_{10})\) reflects the out-of-phase components. }
\label{SolnReduT}
\end{figure*}
\end{widetext}

When performing a dimensional analysis in the context of the differential equation given in Eq.~\eqref{sdot}, the goal is to express the equation in a dimensionless form by introducing a reduced (scaled) time $t' = \frac{t}{\sigma} $. Here \( \sigma \) is a characteristic timescale of the system, often related to the width of a pulse or the inverse of the Rabi frequency. Using the chain rule, the time derivative in the new dimensionless time  $t'$  becomes:
\begin{equation}
\begin{aligned}
\frac{d\mathbf{S}}{dt} = \frac{d\mathbf{S}}{dt'} \cdot \frac{dt'}{dt} = \frac{1}{\sigma} \frac{d\mathbf{S}}{dt'}
\end{aligned}
\label{eq:sdotdim}
\end{equation}

Substituting and rearranging this into the original differential equation, \eqref{sdot}, its form in terms of reduced time takes the following shape:
\begin{equation}
\begin{aligned}
\frac{d\mathbf{S}}{dt'} = \sigma g(t')\mathbf{S}
\end{aligned}
\label{eq:sdotdimt}
\end{equation}

In this expression, $ t'=\frac{t}{\sigma}$ represents the dimensionless time, and $ g(t')$ is the matrix function now expressed as a function of the reduced time $t'.$ The factor $\sigma$ appears due to the change of variables, scaling the differential equation in a way that reflects the natural timescale of the systems.
%Now, we express the matrix \( g(t) \) in terms of the dimensionless time \( t' \). Since \( g(t) \) depends on \( \Omega(t) \), and typically \( \Omega(t) \) has the form \( \Omega(t) = \Omega_0 \exp\left(-\frac{(t - \tau)^2}{\sigma^2}\right) \), we can rewrite \( \Omega(t) \) as a function of \( t' \):

%\[
%\Omega(t) = \Omega_0 \exp\left(-\frac{(t - \tau)^2}%{\sigma^2}\right) = \Omega_0 \exp\left(-\left(\frac{t - %\tau}{\sigma}\right)^2\right) = \Omega_0 \exp\left(-\left(t' - \frac{\tau}{\sigma}\right)^2\right)
%\]

%This the matrix evaluated at the reduced time. Here, $ g(t')$  is now a matrix that depends on the reduced time $ t' $, and the equation is scaled by the factor $\sigma$, which is the characteristic time scale of the system.

The use of reduced units helps simplify the analysis and makes the results more general, as they no longer depend on the specific value of $\sigma$. This dimensionless form can be particularly useful when comparing systems with different time scales or when making theoretical predictions that are independent of specific physical constants.

The numerical method employed is the Runge-Kutta 4th order (RK4) method, a popular and robust algorithm for solving ordinary differential equations (ODEs) with high accuracy. %This method approximates the solution by evaluating the slope of the function at multiple points within each time step and then taking a weighted average of these slopes to update the solution. 
In this context, we are solving the time evolution of a quantum system represented by the Bloch vector $\mathbf{S}(t) = [S_1(t), S_2(t), S_3(t)]$, which encodes the state of the system in terms of population and coherence. The evolution of this vector is governed by the matrix $g(t)$, which depends on the time-dependent Rabi frequency $\Omega(t)$.

%The initial Condition  $\mathbf{S}_0 = [0, 0, 1]$,  $\mathbf{S}_0 = [0, 0, -1]$, or $\mathbf{S}_0 = [0.5, 0.5, 0]$.  This implies that the system starts with full population in one state and no coherence.

After solving for \( \mathbf{S}(t) \), the density matrix elements \( \rho_{00} \), \( \rho_{11} \), and the coherences \( \rho_{01} \), \( \rho_{10} \) are computed. The population terms \( \rho_{00} \) and \( \rho_{11} \) represent the probability of finding the system in one of the two states, while \( \rho_{01} \) and \( \rho_{10} \) describe the coherence between these states.

These quantities are plotted over time to visualize the dynamics of the system, showing how populations and coherences evolve under the influence of the time-dependent field.

%By using reduced units, the results can be interpreted in a generalized framework, making them applicable to various quantum systems without relying on specific physical constants. The plots generated from the numerical solution offer insights into how the quantum system responds to the external field, revealing behaviors like oscillations in population and changes in coherence, which are fundamental to understanding quantum dynamics.

The numerical solution is visualized in Figure (\ref{SolnReduT}), which illustrates the dynamical behavior of the quantum system under different initial conditions. The plots reveal how the population distribution and coherence evolve when the system is initially prepared in either of the two states. Moreover, the plots generated from the numerical solution offer insights into how the quantum system responds to the external field, revealing behaviors like population transfer and changes in coherence, which are fundamental to understanding quantum dynamics.

In the first scenario (i), shown in ~Figure (\ref{SolnReduT}), where the population is initially prepared in state \(|0\rangle\), we observe that after interacting with the external pulse, a portion of the population is transferred to state \(|1\rangle\). This transfer results in the creation of a superposition between the two states, which in turn generates coherence. The presence of coherence indicates that the quantum system is now in a state where it exhibits phase relationships between \(|0\rangle\) and \(|1\rangle\), a hallmark of quantum interference effects.

In the second scenario (ii), shown in ~Figure (\ref{SolnReduT}), the population is initially prepared in the state \(|1\rangle\). After the interaction with the pulse, the population is again redistributed between the two states. However, in this case, the coherence not only emerges, but also undergoes a sign change. This change in sign reflects a phase shift in the quantum state, demonstrating that the phase relationship between \(|0\rangle\) and \(|1\rangle\) is sensitive to the initial preparation of the system.

It is important to note that in the absence of the external pulse, the populations would have remained in their respective initial states, and no coherence would have been generated. This observation underscores the role of the pulse in driving the system's dynamics and creating the conditions necessary for quantum superposition and coherence.

%In the absence of an external pulse, the system would remain in its initial state, meaning that the populations of the quantum states would not change, and no coherence would be generated. The pulse, typically represented by \(\Omega(t)\), is responsible for inducing transitions between quantum states, creating the necessary conditions for quantum superposition and coherence. Without this driving force, the system stays in its initial configuration, with no off-diagonal elements (coherences) appearing in the density matrix.

However, if the system was initially excited, even in the absence of a pulse, spontaneous emission would occur. %This is a fundamental process in which the excited state decays to a lower-energy state, releasing a photon. 
Nevertheless, for the proposal at hand, it is sufficient to assume that the computation occurs on a timescale much faster than the decay time due to spontaneous emission. %This means that the influence of spontaneous emission can be neglected during the time frame of interest, allowing for a clean analysis of the pulse-induced dynamics without the need to consider decay effects.

%This approach simplifies the modeling, focusing on the coherent dynamics driven by the pulse and omitting the effects of spontaneous emission, which would become relevant only on longer timescales. Consequently, the system remains in a controlled quantum superposition, and coherence is maintained during the computation, as the external pulse drives transitions between the quantum states.

These results highlight how the dynamics of an atomic system can resemble the behavior of a finite-state machine. Just as in a finite-state machine, where the output depends not only on the input but also on the current state of the system, the evolution of the quantum system is determined by both the applied pulse (the input) and the initial quantum state. This analogy emphasizes the importance of both the external driving force and the system's initial conditions in determining the final state of the system.

In the next section, we will explore how the dynamics of a quantum system can emulate a finite-state machine, specifically focusing on its behavior as a parity checker. By analyzing how the quantum system transitions between states and how coherence evolves in response to the pulse, we can demonstrate how it mimics the behavior of a parity checker. This analogy highlights the system's ability to process input (the pulse) and provide output (coherence and population) based on its initial conditions, similar to an FSM evaluating input sequences.

In summary, the next section will detail how the quantum system's dynamics reflect the principles of a finite-state machine, specifically illustrating its function as a parity checker. This comparison underscores the versatility of quantum systems in modeling and simulating classical computational tasks.
%%%%%%%%%%%%%%%%%%%%%%%%%%%%%%%%%%%%%%%%%%%%%%%%%%%%%%%%%%%%%%

%%%%%%%%%%%%%%%%%%%%%%%%%%%%%%%%%%%%%%%%%%%%%%%%%%%%%%%%%%%%%
\section{Parity Checker}
\label{sec: parity}
%%%%%%%%%%%%%%%%%%%%%%%%%%%%%%%%%%%%%%%%%%%%%%%%%%%%%%%%%%%%%%
In this discussion, we examine how the dynamics of a two-level system can be harnessed to emulate the functionality of a Parity Checker. Initially, the system is set to the ground state $|0\rangle$. By applying a laser pulse, we can transfer the population to the excited state $|1\rangle $. Depending on the shape and duration of the pulse, we can also induce coherence between these two states.

To utilize this system's dynamics for logic operations, we first need to define our input and output variables. The inputs consist of two factors: the state of the laser pulse (whether it is \textbf{ON} or \textbf{OFF}) and the initial state of the system (whether it is in $|0\rangle$ or $|1\rangle$ ). The outputs are determined by measuring the coherences between the states $|0\rangle$ and $|1\rangle$, as well as the final populations of these states after the interaction with the laser pulse.

The process involves assigning logic values to these outputs based on their measurements. Specifically, if the population of a state $|m\rangle$ is $\geq 0.6 $ following the laser pulse interaction, a logic value of \textbf{1} is assigned. In contrast, if the population is below $0.6$, it is assigned a logic value of \textbf{0}. For coherences, a logic value of \textbf{1} is assigned if the absolute value of the coherence is $\geq 0.5 $, and a logic value of \textbf{0} is assigned if it is less than $0.5$. This assignment of logic values based on measured outputs effectively maps the dynamics of the two-level system to the behavior of a Parity Checker, as detailed in the truth table of Table~(\ref{table:parity}).

%\begin{widetext}
\begin{table}[htp]
\centering
\begin{tabular}{ | l| l| l| l | }
\hline
&  &  &   \\
 \textbf{Pulse} & \textbf{Initial } & \textbf{Final}  & \textbf{Coherence } \\
      $\Omega(t)$                & \textbf{ State} &  \textbf{ State}&   \\
\hline
0  & 0 & 0 & 0   \\
0  & 1 &  1 & 0 \\
1  & 0 &  1 & 1   \\
1  & 1 & 0 &  1   \\
\hline
%$\Omega(t)$   &$s(t_0)$&    & \textbf{$s(t)$}   \\
%\hline
\textbf{PS}   & \textbf{PI} &  \textbf{NS}  & \textbf{PO}\\
$\boldsymbol{s(t)} $  & $\boldsymbol{x(t)}$  &  $\boldsymbol{s(t+1)}$   & $\boldsymbol{z(t)}$  \\
\hline
\end{tabular}
\caption{Truth table for Parity checker, where PS, NS, PI, and PO stand for the present state, the next state, present input, and present output respectively. }
\label{table:parity}
\end{table}
%\end{widetext}

In our logic assignment scheme, the logic value for the pulse is defined as follows: It is \textbf{ 0} when the pulse is \textbf{OFF} and \textbf{1} when the pulse is \textbf{ON}. For the initial state, the logic values are based on the population states: a state where all populations are in \( |0\rangle \) is assigned a logic value of \textbf{0}, while a state where all populations are in \( |1\rangle \) is assigned a logic value of \textbf{1}. Since there are two inputs—pulse state and initial state—and each can be either \textbf{0} or \textbf{1}, there are \(2^2 = 4\) possible combinations of these inputs, as illustrated in the truth table of Table~(\ref{table:parity}).

The final logic value for the populations in the state after interaction with the laser pulse is assigned based on the following criteria: if the population of a state \( |m\rangle \) is \( \geq 0.6 \), it is assigned a logic value of \textbf{1}. If the population is below \(0.6\), it is assigned a logic value of \textbf{0}. Thus, a population of \( \geq 0.6 \) corresponds to a logic value of \textbf{1}, while a lower population corresponds to \textbf{0}. For coherences, a logic value of \textbf{1} is assigned if the absolute value of the coherence is \( \geq 0.5 \), and a logic value of \textbf{0} is assigned if the absolute value is less than \(0.5\).
%In our logic assignment scheme, the pulse's logic value is straightforward: it is \textbf{0} when the pulse is \textbf{OFF} and \textbf{1} when the pulse is \textbf{ON}. For the initial state, the logic values are assigned based on the population states: a state where all populations are in \( |0\rangle \) is assigned a logic value of \textbf{0}, while a state where all populations are in \( |1\rangle \) is assigned a logic value of \textbf{1}. With two inputs—pulse state and initial state—each of which can be either \textbf{0} or \textbf{1}, there are a total of \(2^2 = 4\) possible input combinations. The final logic assignment for the final state populations is determined as follows: if the population of a state \( |m\rangle \) is \( \geq 0.6 \) after the system interacts with the laser pulse, it is assigned a logic value of \textbf{1}. If the population is below \(0.6\), the assigned logic value is \textbf{0}. Thus, a population of \( \geq 0.6 \) corresponds to a logic value of \textbf{1}, while a lower population corresponds to \textbf{0}. Similarly, for coherences, a logic value of \textbf{1} is assigned if the absolute value of the coherence is \( \geq 0.5 \), and a logic value of \textbf{0} is assigned if the absolute value is less than \(0.5\).

In general, the parity of a bit string refers to the evenness or oddness of the total number of 1 bit contained in the bit string. Parity checking is commonly adopted in digital circuits. A string of bits has an 'even parity' if the number of 1-bits in the string is even; otherwise, it has odd parity \cite{kohavi2010switching}.

The superevolution operator that is obtained using the Sylvester formula propagates the system from the initial time $t_0$ to some final time $t$, coupled with the profile of the laser pulse used, and the dynamics are seen to follow operations that are similar to those of the finite-state machine operation.  

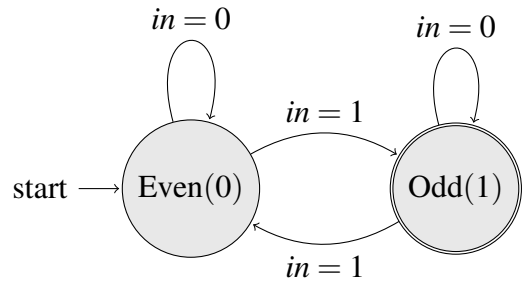
\begin{figure}[htp]
\centering
\begin{tikzpicture}[shorten >=1pt,node distance=3.5cm,on grid, auto]
  \tikzstyle{every state}=[fill={rgb:black,1;white,10}]

    \node[state,initial]   (q_1)                    {$\mathrm{Even(0)}$};
    \node[state,accepting] (q_2)  [right of=q_1]    {$\mathrm{Odd(1)}$};
    %\node[state]           (q_3)  [right of=q_2]    {$q_3$};

    \path[->]
    (q_1) edge [loop above] node {$in=0$}    (   )
          edge [bend left]  node {$in=1$}    (q_2)
    (q_2) edge [bend left]  node {$in=1$}    (q_1)
          edge [loop above] node {$in=0$}    (   );
    %(q_3) edge [bend left]  node {0,1}  (q_2);
\end{tikzpicture}
 \caption{ State diagram of two state parity checker} \label{fig:parity}
 \end{figure}

As shown in the schematic representation in Fig.~\ref{fig:parity}, the circuit operates with two states --- either even or odd --- corresponding to the pulse being \textbf{OFF} or \textbf{ON}, respectively, accepts a series of bits in serial and outputs \textbf{0} if the parity so far is even and outputs \textbf{1} if odd. A parity checker bit is added to bits of the string to ensure the total number of 1-bits is even or odd. We represent these two states as circles in Fig.~\ref{fig:parity}, at any given moment, the machine is in either of the states, which we encode \textbf{0} for the even state, and \textbf{1} for the odd state. We have an input bit (\textbf{In=0} or \textbf{1} in the figure: —corresponding to the initial states $|0\rangle$ and $|1\rangle$, respectively) that indicates the machine that transitions. Because we have one input bit,  there are two possible transition directions (shown as \textbf{0}, \textbf{1} in the figure), which show the direction to go to the machine. The output corresponding to the state is produced once the machine makes the transition, and the output can remain in the same state or move into a different state. 

You can use either a state table or a state diagram to describe the relationships between the input symbol $x(t)$, the present state $s(t)$, the output symbol $z(t)$, and the next state $s(t+1)$. For each combination of the input symbol and the present state, the corresponding entry provides information about the output that will be produced and the next state to which the machine will go. Each circle in the state diagram represents the state of the machine (cf. Fig.~\ref{fig:parity}). Each circle emanates directed arcs, indicating state transitions caused by the input. The directed arc is labeled by the input symbol that causes the transition. Because both the state table and the state diagram contain the same information, the choice between the two representations is a matter of convenience \cite{kohavi2010switching}, and we use the state diagram, Fig.~(~\ref{fig:parity}), to show the relationship between the input, initial states, and corresponding transitions of the parity checker.

Note that if the state of the machine is even (corresponding to logic value \textbf{0}), then input string \textbf{0} dictates to stay in the same state outputting \textbf{0} (i.e. even); similarly input string \textbf{1} means changing state to odd (corresponding to logic value \textbf{1}) outputting \textbf{1} (i.e. odd). In contrast, if the state of the machine is odd (corresponding to the logic value \textbf{1}), then the input string \textbf{0} means staying in the same state, thereby putting \textbf{1} (i.e., odd), and the input string \textbf{1} means changing the state to even (corresponding to the logic value \textbf{0}) and thus putting \textbf{0} (i.e., even). 

Considering this logic machine, it is important to note that the dynamics, logic assignments, and variables used are the same as those discussed in~\cite{hailu2019su2}. The present state of the logic machine, the state of even or odd parity, is encoded as the state of the pulse being \textbf{OFF} or \textbf{ON}, respectively. The state of the quantum system, that is, the physical system, is encoded to represent the input of the parity checker, which is \textbf{1} if the population is initially prepared to be in the state $|1\rangle$ otherwise it is \textbf{0}. In this context, the term initial state refers to the state of the machine before the application of the input, whereas the state (of the machine) after the application of the input is called the final state. The next state of the machine is determined by the final state of the two-level system, that is, after interacting with the laser. If the system ends in state $|1\rangle$, it corresponds to the logic value \textbf{1}; otherwise, it corresponds to \textbf{0}. Finally, the output of the parity checker is encoded by the presence or absence of coherence.
%\graphicspath{{Images//}}
\begin{figure}[htbp]
\begin{center}
%(a) \includegraphics[width=2.0 in]{Images/2level.png}
 \includegraphics[width=3.5 in]{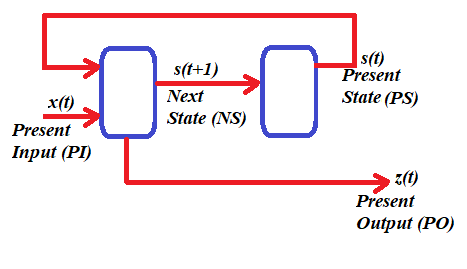}
\caption{(Color online) Finite State machine where the outputs not only depend on the input but also rely on the state it is in.}
\label{FSM}
\end{center}
\end{figure}

In a finite-state machine (FSM), the key components are typically denoted as follows. The input, represented as $x(t)$, denotes the external signals or stimuli that influence the behavior and state transitions of the FSM. The output, denoted as $z(t)$, represents the result or action produced by the FSM based on its current state and the input it receives. Finally, the next state, often denoted $s(t+1)$, is the state to which the FSM transitions after processing the input. The next state is determined by the current state, $s(t)$, and the input according to the FSM's transition function. %In FSM diagrams, states are usually depicted as nodes or circles, transitions as arrows between these nodes, and inputs and outputs are indicated through labels on the arrows or annotations associated with the states or transitions.

As a final note, we highlight that the proposed operation is sensitive to several parameters, particularly detuning and environmental effects. The inclusion of noise, for example, will degrade coherence, rendering the proposed implementation of the finite-state machine unfeasible. %It is important to note that when noise is taken into account, these observables may disappear. 
However, the timescale of the proposed machine is sufficiently rapid to complete the computation before the stored information is lost. Additionally, employing a detuning value significantly far from resonance complicates the analytical solution, as the Hamiltonian no longer commutes with itself at different times.
%%%%%%%%%%%%%%%%%%%%%%%%%%%%%%%%%%%%%%%%%%%%%%%%%%%%%%%%%%%%%%
\section{Scalability}
\label{sec: scalability}
%%%%%%%%%%%%%%%%%%%%%%%%%%%%%%%%%%%%%%%%%%%%%%%%%%%%%%%%%%%%%%
%One of the primary advantages of using rare earth ions is their sharp optical transitions and long decoherence times, making them ideal for coherent adiabatic processes. Importantly, all three ground states of these ions can couple with all three excited states, allowing for nine potential optical transitions.

Rare-earth ions embedded in solid-state hosts, such as Y$_2$SiO$_5$, present a scalable platform for quantum information processing. Their sharp optical transitions and long coherence times are critical for enabling coherent adiabatic processes, which form the backbone of many quantum operations. These systems exhibit exceptional stability as a result of the robust environment provided by the host crystals, minimizing decoherence and enhancing scalability. 

For instance, europium-doped Y$_2$SiO$_5$ has demonstrated nuclear spin coherence times of up to six hours, highlighting the potential for long-term preservation of the quantum state \cite{ZhongNature}. This level of coherence enables high-fidelity quantum operations necessary for memory and computational tasks, making such systems attractive for large-scale quantum networks.

Rare-earth-doped crystals also facilitate the transition from two-level systems to N-level systems, broadening computational possibilities. The multilevel coupling capabilities allow for scalable architectures capable of hosting more complex algorithms. Strong light-matter coupling has been observed in these systems, particularly in the high-cooperativity regime, where interactions between collective spins and optical modes are enhanced \cite{AlexanderRareEarth}. Such properties are critical for achieving the scalability required for advanced quantum technologies.

Furthermore, advances in electromagnetically induced transparency (EIT) have enabled optical storage times exceeding one second in solid-state systems like Pr$^{3+}$:Y$_2$SiO$_5$ \cite{LongdellStoppedLight}. These findings reinforce the practicality of these systems for scalable quantum operations. The ability to mitigate spectral diffusion through optimized environmental conditions further enhances the scalability of rare-earth-ion systems, paving the way for their integration into larger quantum architectures.

These features collectively position rare-earth-ion-based quantum systems as a cornerstone for developing scalable quantum technologies, from small-scale implementations to complex multilevel operations.

Now the question is whether we can broaden the equation of motion to an N-level system, to this end, we revisit our calculation of Section\ref{sec:thesystem} and certainly! To extend the derivation to an \(N\)-level system, we need to account for the fact that the algebra in an \(N\)-level quantum system is described by \(N^2 - 1\) generators, instead of just the three Pauli matrices. These generators correspond to the Lie algebra \(su(N)\), and can be expressed in terms of the matrix elements \( \hat{O}_{mn} = |m\rangle\langle n| \) for \(m \neq n\).

\subsection{Extending to \(N\)-Level Systems}

For a quantum system with \(N\) levels, we generalize the Pauli matrices at two levels to a set of \(N^2 - 1\) generators that form the algebra \(su(N)\). These generators can be expressed as linear combinations of projector operators \(\hat{O}_{mn} = |m\rangle \langle n|\). The specific construction of these generators is more complex but follows a structure similar to the Pauli matrices. In the two-level case, we had three generators (\(\hat{s}_1, \hat{s}_2, \hat{s}_3\)) corresponding to the \(SU(2)\) algebra. For systems of the level \(N\), there are generators \(N^2 - 1\), which are commonly denoted as \(\hat{s}_j\) where \(j = 1, 2, ..., N^2 - 1\).

In terms of these generators, we can expand the density matrix and Hamiltonian in an \(N\)-level system as follows, following similar steps as in the two-level case:
\begin{equation}
\begin{aligned}
\hat{\rho}(t) = \frac{\hat{I}}{N} + \frac{1}{2} \sum_{j=1}^{N^2 - 1} S_j(t) \hat{s}_j,
\end{aligned}
\label{eq:rhoexpN}
\end{equation}
and
\begin{equation}
\begin{aligned}
\hat{H}(t) = \frac{\hbar}{2} \left[\left(\sum_{k=1}^N \omega_k \right) \hat{I} + \sum_{j=1}^{N^2 - 1} h_j(t) \hat{s}_j \right],
\end{aligned}
\label{eq:HamexpN}
\end{equation}
where \(\hbar \omega_k\) represents the energy of the \(k\)-th level, and \(\hat{I}\) is the identity operator. The coefficients \(S_j(t)\) and \(h_j(t)\) are computed as follows:

\begin{equation}
\begin{aligned}
S_j(t) = \text{Tr}\left(\hat{\rho}(t) \hat{s}_j\right), \quad h_j(t) = \frac{1}{\hbar} \text{Tr}\left(\hat{H}(t) \hat{s}_j\right).
\end{aligned}
\label{eq:GenexpN}
\end{equation}

%### Properties of the Generators

The generators \(\hat{s}_j\) in a \(N\) level system have properties similar to the Pauli matrices in the two-level case. Specifically, they satisfy:

\begin{equation}
\begin{aligned}
\text{Tr}(\hat{s}_i \hat{s}_j) = 2 \delta_{ij}, \quad [\hat{s}_i, \hat{s}_j] = 2i \sum_{k=1}^{N^2 - 1} \epsilon_{ijk} \hat{s}_k,
\end{aligned}
\label{eq:TrN}
\end{equation}

where \(\delta_{ij}\) is the Kronecker delta and \(\epsilon_{ijk}\) is the Levi-Civita symbol, which is antisymmetric on the three indices.

%%%### Equation of Motion for the Coherence Vector

Following a similar procedure to the two-level case \cite{Hioe1981, AlhassidLevine}, we now define the coherence vector \(\vec{S}(t)\) as an \(N^2 - 1\) dimensional vector whose components are the expectation values \(S_j(t)\) of the generators \(\hat{s}_j\), given by $\vec{S}(t) = \left(S_1(t), S_2(t), ..., S_{N^2 - 1}(t)\right)^T$.

The evolution of this vector is governed by the following equation of motion:

\begin{equation}
\begin{aligned}
\frac{d}{dt} \vec{S}(t) = g(t) \vec{S}(t), \quad \vec{S}(t_0) = \vec{S}_0,
\end{aligned}
\label{eq:sdotN}
\end{equation}

where \(g(t)\) represents the time-dependent interaction coefficient that governs the dynamics. This equation describes the time evolution of the coherence vector in the presence of a perturbation, analogous to the two-level case, but with a much larger set of generators corresponding to the dimensions \(N^2 - 1\) of the \(su(N)\) algebra.

The solution of equation \eqref{eq:sdotN} can be formally written as 
\begin{equation}
\begin{aligned}
\vec S\left(t\right)=&Q\big(t,t_0\big)\vec S\left(t_0\right)
\end{aligned}
\label{transform}
\end{equation}
where $\vec S$ being an $(N^2 - 1)$-dimensional column vector whose components are the average values of the observables, and the super-evolution matrix $Q$ an $(N^2 - 1)\times (N^2 - 1)$ matrix. 
% Summary of Key Points

In an \(N\)-level system, the dynamics are governed by \(N^2 - 1\) generators corresponding to the \(su(N)\) algebra, extending the Pauli matrices from the two-level case. The density matrix and the Hamiltonian are expanded in terms of these generators, and the coefficients are computed from the expectation values of the generators.
The equation of motion for the coherence vector follows a form similar to the two-level case, with the state vector evolving according to a time-dependent coefficient \(g(t)\).

This section extends the analysis from two-level systems to \(N\)-level systems, incorporating the appropriate number of generators (\(N^2 - 1\)) and maintaining the structure of the equations. 

%%%%%%%%%%%%%%%%%%%%%%%%%%%%%%%%%%%%%%%%%%%%%%%%%%%%%%%%%%%%%%%%
\section{Linear Sequential Machine}
\label{sec: LSM} 
%%%%%%%%%%%%%%%%%%%%%%%%%%%%%%%%%%%%%%%%%%%%%%%%%%%%%%%%%%%%%%%%
In this section, we explore the relationship between the superevolution matrix, discussed in the previous section, and linear sequential machines (LSMs). A linear sequential machine is defined by the way the output and the next state of the machine are determined based on the current input and the present state of the machine \cite{surveyFSM,kohavi2010switching, remaclepar}. LSMs are typically described by equations of the form:

\begin{equation}
\begin{aligned}
s(t + 1) &= A s(t) + B u(t) \\
y(t) &= C s(t) + D u(t)
\end{aligned}
\label{eqLSM}
\end{equation}

Here, the vector \(s(t)\) represents the vector of the present state, whose elements are the state variables, and \(s(t+1)\) is the vector of the next state. The vector \(u(t)\) denotes the input vector and \(y(t)\) is the output vector. The matrices \(A\), \(B\), \(C\), and \(D\) are the characterizing matrices of the system. If \(D \neq 0\), the machine is classified as a Mealy machine; otherwise, it is a Moore machine.

The first equation in \eqref{eqLSM} illustrates that the next state of the machine depends on both the current input and the present state of the machine. This behavior mirrors the superevolution process described earlier in Eq.\eqref{transform}, where the observable vector evolves from its initial value at time \(t_0\) to its value at time \(t\). 

The output \(y(t)\) of the machine could, though not necessarily, be the same as the state of the machine. In such a case, we would have \(C = I\) and \(D = 0\) (with \(I\) being the identity matrix). In all machines considered in this paper, the output does not depend directly on the input (that is, \(D = 0\)). Instead, the output is typically interpreted as an expectation value of one or more observables, which means that the structure of the matrix \(C\) depends on the measured observables.

At this point, it is important to distinguish between two scenarios regarding the observable vector at time \(t+1\): whether it occurs shortly after time \(t\) or after a longer duration. We discuss both cases in the following.

\subsection{Short Time Evolution} 

If \(t+1\) represents a time shortly after \(t\), the state change is expected to be small. In this case, the observable vector \(S\), as discussed in Eq.\eqref{eq:sdotN}, can be approximated using a first-order Taylor expansion:

\begin{equation}
\begin{aligned}
S(t + \delta t) &\approx S(t) + \left( \frac{dS(t)}{dt} \right) \delta t \\
&\approx S(t) + g(t) S(t) \delta t
\end{aligned}
\label{LSMtaylor}
\end{equation}

Here, \(\delta t\) is small compared to the timescale of the physical process. This is consistent with the concept of Lie algebra, which is derived from linearizing Lie groups. Linearization, which is essentially a Taylor series expansion around the identity, generates new operators that form the Lie algebra \cite{Gilmore2006}. Since the components of the observable vector are expectation values of the group generators, the evolution of the observable vector over short time intervals (that is \(t + \delta t\)) is governed by the Lie algebra. As indicated in Eq.\eqref{LSMtaylor}, the next state of the observable vector \(S(t + \delta t)\) depends on both its current state \(S(t)\) and the applied input \(g(t)\).

\subsection{Long Time Evolution}

If \(t + 1\) represents a significantly longer time after \(t\), the state change may be substantial. These types of dynamics can be analyzed using a superevolution matrix. The super-evolution matrix propagates the observable vector from its initial value at \(t_0\) to its value at time \(t\). 

The observable vector is far from its initial value for long-time evolution due to perturbations. Therefore, using the Lie group framework is appropriate, as we are far from the identity. The super-evolution matrix \(Q(t, t_0)\) can be related to the matrix \(A\) in Eq.\eqref{eqLSM} by defining an appropriate timescale for the machine's evolution. This allows us to model the long-term behavior of the system and the evolution of its observable vector.
Mitigating and controlling noise in systems modeled by equation \eqref{eq:sdotN} or similar quantum systems' dynamics involves a combination of tailored techniques to address specific noise sources and system dynamics. An effective approach is to enhance the coherence times of quantum states through environmental isolation and precise control of system parameters. By minimizing exposure to external disturbances, such as electromagnetic interference or thermal fluctuations, noise can be significantly suppressed. Additionally, employing error correction methods, like Quantum Error Correction (QEC), helps manage and mitigate noise by encoding information in a way that can tolerate noise-induced errors. Fault-tolerant techniques are essential for ensuring the accuracy of quantum gates and measurements, even in the presence of noise, as they rely on redundancy and real-time error detection and correction. Optimal control theory plays a crucial role in noise mitigation by enabling precise manipulation of system parameters to counteract noise effects. This may involve tailor-made pulse sequences and feedback mechanisms to dynamically adjust quantum operations. Furthermore, integrating machine learning approaches can enhance noise mitigation by predicting and compensating for errors based on historical data, thereby improving the system's resilience to noise over time. By combining strategies such as isolation, error correction, and precise control, quantum systems modeled by Equation \eqref{eq:sdotN} can effectively manage and control noise, leading to more accurate and reliable quantum computations and simulations.
%%%%%%%%%%%%%%%%%%%%%%%%%%%%%%%%%%%%%%%%%%%%%%%%%%%%%%%%%%%%%%%%%%%%%%%%%%%%%%%%%%%
\section{Comparison Between the Proposed Finite State Machine and Traditional Logic Circuits} 
\label{sec:compare}  
%%%%%%%%%%%%%%%%%%%%%%%%%%%%%%%%%%%%%%%%%%%%%%%%%%%%%%%%%%%%%%%%%%%%%%%%%%%%%%%%%%%
We intend to conduct a detailed exploration of the power efficiency and computational complexity of the proposed machine in future work. For now, however, we will briefly compare its computational complexity to that of traditional logic systems, which are grounded in classical computational paradigms. The proposed finite-state machine leverages the dynamics of atomic systems, offering a unique approach to computation that fundamentally differs from traditional silicon-based circuits. By utilizing quantum observables, which can exist in superpositions and exhibit population dynamics, this logic machine harnesses the distinct properties of quantum mechanics to enable parallelism, nonlocality, and, consequently, faster computation. Unlike classical logic circuits built on Boolean algebra and classical bits, the proposed machine exploits the discreteness of quantum states, offering a richer computational framework that extends beyond Boolean logic.

%To provide a mathematical proof for the statement that a classical computer may require an exponential number of operations to solve a complex problem, while the proposed quantum machine could achieve polynomial solutions due to its ability to manage a broader spectrum of states concurrently, we can analyze the growth rates of computational complexity.

\subsection{Classical Computational Complexity}

Let \( N \) represent the size of a problem that a classical computer must solve. The time complexity to solve this problem, \( T_{\text{classical}}(N) \), is typically expressed as a function of \( N \). For many complex problems, especially combinatorial ones, the complexity of time grows exponentially with \( N \). This can be represented as follows.

\begin{equation}
\begin{aligned}
T_{\text{classical}}(N) = 2^N
\end{aligned}
\label{tclassical}
\end{equation}

This exponential growth indicates that the number of operations required increases very rapidly with the size of the problem, making it infeasible for large values \( N \) \cite{therycomptn}.

\subsection{Computational Complexity of the Proposed Machine} 
On the other hand, the proposed machine takes advantage of both the discreteness of quantum states and the principle of superposition to achieve computational efficiency. By exploiting superposition, the system can represent and process multiple quantum states simultaneously, allowing it to explore a broader state space in fewer steps. If the machine operates with discrete quantum states $M$ and uses superposition, the number of accessible states grows exponentially as $d^M$, where $d$ is the dimensionality of each state. However, the actual computational complexity does not scale exponentially because the machine employs efficient quantum transitions and algorithms. For example, in a quantum state space with \( d \) possible states, a quantum computer can explore all \( d \) states simultaneously. If a quantum machine has \( M \) qubits, the number of possible quantum states is \( 2^M \). However, it does not require \( 2^M \) operations to perform computations, thanks to quantum parallelism and efficient quantum algorithms \cite{MichaelChuang2010}.

The key advantage of quantum computing is its ability to solve problems more efficiently through polynomial-time algorithms. For example, Grover’s algorithm provides a quadratic speed-up, solving search problems in \( O(\sqrt{N}) \) rather than \( O(N) \) \cite{Grover}, and Shor’s algorithm efficiently factorizes integers in polynomial time \cite{PShor}. Thus, for many computational tasks, the complexity can be expressed as:

\begin{equation}
\begin{aligned}
T_{\text{proposed}}(N) = N^k, \quad k < 2
\end{aligned}
\label{tqm}
\end{equation}
Here, \( k \) is a constant less than 2, indicating polynomial growth. This polynomial growth contrasts with the exponential growth seen in classical computation \cite{Preskill2018quantumcomputingin}.

Furthermore, as the dynamics are optically addressed, the proposed finite-state machine demonstrates the potential for computations at significantly higher speeds compared to classical counterparts, which typically operate on electrical signals. In addition, the discrete nature of the system's domains facilitates massive parallelism or optimization tasks. Quantum mechanics allows these machines to leverage superposition and entanglement, enabling simultaneous exploration of multiple states. This capability allows for efficient handling of exponentially larger state spaces, potentially reducing computational time exponentially. For example, while a classical computer may require an exponential number of operations to solve a complex problem, the proposed machine could achieve polynomial solutions because of its ability to manage a broader spectrum of states concurrently.

Moreover, the finite-state machine excels in precision tasks involving quantum-state manipulation, probabilistic outcomes, and interference effects. Unlike classical circuits, which operate deterministically, quantum machines process information in probabilistic terms dictated by the principles of superposition and entanglement. This allows for highly accurate solutions to specific problems, though it may come at the cost of added complexity in error correction and maintaining coherence, as quantum states are extremely susceptible to noise and decoherence. Traditional circuits provide deterministic precision, supported by decades of advances in error detection and correction methods, ensuring high reliability even in extensive computational operations.

The proposed finite-state machine could be implemented using specialized hardware distinct from traditional silicon circuits. This type of machine may require components such as superconducting qubits, trapped ions, or photonic setups. A promising candidate, for example, is quantum dots as discussed in \cite{FreschQDT, klein2010ternary, klein2007transcending}.
Therefore, the proposed quantum finite-state machine leverages quantum mechanics to manage a broader spectrum of states concurrently, reducing the number of operations required to solve complex problems compared to classical systems. Mathematically, this manifests itself as polynomial complexity rather than exponential complexity, providing a significant computational advantage for specific problem domains \cite{MichaelChuang2010}.
In summary, the proposed finite state machine based on atomic dynamics presents a powerful paradigm for specific computational challenges, particularly where parallelism and exponential state management are essential. However, its current challenges in terms of energy consumption, error rates, and hardware complexity highlight the obstacles it faces compared to traditional silicon-based circuits. While quantum computing holds transformative potential for fields like cryptography, materials science, and drug discovery, classical circuits continue to serve as the cornerstone of general computing infrastructure for a wide array of applications.
%%%%%%%%%%%%%%%%%%%%%%%%%%%%%%%%%%%%%%%%%%%%%%%%%%%%%%%%%%%%%%%%%%%%
\section{Future Direction}
\label{sec:future}
%%%%%%%%%%%%%%%%%%%%%%%%%%%%%%%%%%%%%%%%%%%%%%%%%%%%%%%%%%%%%%%%%%%%
Mitigating and controlling noise in systems modeled as quantum systems, discussed in this paper, involves a combination of tailored techniques to address specific noise sources and system dynamics. An effective approach is to enhance the coherence times of quantum states through environmental isolation and precise control of system parameters, such as praseodymium ions doped in yttrium silicate as discussed in section(\ref{sec:thePhyssystem}), which is crucial for robust quantum operations. These systems inherently exhibit significantly longer coherence times because of their weak coupling to environmental disturbances and the long lifetimes of their excited states \cite{LongdellStoppedLight}. These properties differentiate them from typical quantum systems and make them particularly advantageous for applications in quantum information processing and simulations. Despite these benefits, noise mitigation remains an essential challenge that requires both theoretical and practical advancements.

Another promising approach to mitigating noise involves dynamical decoupling, a technique designed to reduce the effects of dephasing and other forms of noise by periodically reversing the dynamics of the system\cite{DynamicalSupression, QMBitAlive}. This is achieved through the application of a time-dependent control field, where the amplitude alternates in sign at regular intervals. The control field \(\Omega(t)\), when modified to implement dynamical decoupling, becomes \(\Omega_{\text{DD}}(t) = (-1)^n \Omega(t)\), where \(t \in [nT, (n+1)T)\) and \(T\) is the duration of each interval. The toggling frame created by this periodic modulation ensures that the system averages out low-frequency noise over time, effectively decoupling it from the surrounding environment. 

The influence of this control strategy can be analyzed by examining the effective Hamiltonian that governs the system's dynamics under dynamical decoupling. This Hamiltonian, averaged over a complete cycle of the control field, reflects the reduced impact of noise. The evolution of the density matrix \(\rho(t)\) or equivalent observable vector components under this modified Hamiltonian shows preserved coherence, even in the presence of environmental fluctuations. This preservation of quantum states demonstrates the effectiveness of dynamical decoupling as a noise suppression method.

The implementation of dynamical decoupling and other advanced noise mitigation strategies, such as error correction \cite{PShorDecoh} and optimal control, requires detailed calculations and experimental validation tailored to the specific dynamics of rare-earth-ion-doped systems. Future work should focus on quantifying the impact of these techniques, optimizing control sequences, and integrating feedback mechanisms to further enhance the system's noise resilience. Additionally, the exploration of hybrid approaches that combine dynamical decoupling with error correction could unlock new pathways for achieving fault-tolerant quantum operations in these systems.

Using the naturally extended coherence times of rare-earth ion-doped systems and incorporating advanced noise mitigation techniques \cite{PRXQuantumNoise,KHANEJA2005296}, it is possible to create a highly robust platform for quantum computing and simulations. The potential for these systems to maintain coherence over extended durations highlights their suitability for scalable and reliable quantum technologies.
\section{Conclusion}
\label{sec: conclusion}
%%%%%%%%%%%%%%%%%%%%%%%%%%%%%%%%%%%%%%%%%%%%%%%%%%%%%%%%%%%%%%
In conclusion, we have shown that a Hamiltonian governing the dynamics of a quantum system can be effectively represented as a linear combination of elements from the \(SU(2)\) group, which serves as a fundamental group for describing two-level quantum systems. By selecting an appropriate pulse profile and detuning, we can design an evolution operator that mimics a parity checker, a class of finite-state machine (FSM). The evolution matrix, which governs the time evolution of the system, is constructed to map the system from one state to another while maintaining the fidelity of the information it carries.

In this approach, the system's observables, population, and coherence (both diagonal and off-diagonal elements of the density matrix) play an essential role in encoding and processing information. The diagonal elements, representing populations of quantum states, provide information about the system's current state, while the off-diagonal coherence terms capture the quantum superposition and coherence between states. Together, they act as the "information bits" of the system, reflecting the encoded quantum information.

When the evolution matrix acts on the initial state of the system, it transforms the system so that the output state reveals the properties of the initial state, preserving the encoded information. This is a key feature of the proposed model, where the FSM logic can be seen in the transformation of quantum states under controlled operations. This process parallels the behavior of classical FSMs, which track and transition between discrete states in response to inputs, with the key distinction being that the quantum system exhibits the inherent features of superposition and entanglement.

Moreover, the finite-state machine model we have outlined here is not restricted to a two-level system. It can be extended to an \(N\)-level system, where the quantum state space is enlarged, and more complex operations can be encoded. This scalability makes the model applicable to a wider range of quantum systems, including those with higher dimensionalities, such as multi-level quantum systems. %or systems involving quantum bits (qubits) arranged in higher-order entanglements. The ability to generalize the finite state machine model to larger systems opens new possibilities for quantum computation, simulation, and information processing, as it provides a robust framework for manipulating complex quantum states in a controlled manner. By employing this framework, we can begin to explore new quantum algorithms and error-correction methods that exploit the quantum mechanical advantages of coherence and entanglement while mitigating the effects of noise and decoherence.
%%%%%%%%%%%%%%%%%%%%%%%%%%%%%%%%%%%%%%%%%%%%%%%%%%%%%%%%%%%%%%

\section*{References}

\bibliography{mybibfile}

\end{document}